# Example of lumped parameter modeling of a quantum optics circuit and proposed decisive test of time-symmetric physics

Paul J. Werbos*
ECCS Division, National Science Foundation
Retired, 2015

**Abstract**

Is it possible for a simple lumped parameter model of a circuit to yield correct quantum mechanical predictions of its behavior, when there is quantum entanglement between components of that circuit? This paper shows that it is possible in a simple but important example – the circuit of the original Bell's Theorem experiments, for ideal polarizers. Correct predictions emerge from two alternative simple models, based on classical Markov Random Fields (MRF) across space-time, which are local and realistic and symmetric with respect to time. Exact agreement with quantum mechanics does not violate Bell's Theorem itself, because the interplay between initial and final outcomes in these calculations does not meet the classical definition of time-forwards causality. Both models raise interesting questions for future research. The final section discusses several possible directions for following up on these results, both in lumped system modeling and in more formal and general approaches. It describes how a new all-angles triphoton experiment, not yet performed, would decisively tell us which is true, either the time-symmetric MRF models or the time-forwards collapse model assumed in the usual measurement formalism of Copenhagen quantum mechanics. A new appendix calculates what the Copenhagen measurement formalism predicts for the triphoton experiment, and provides a continuous time formulation of the usual collapse of the wave function without metaphysical observers.

## 1 Introduction

Lumped parameter circuit models use just a handful of variables to describe the state of each component in a circuit. The components are connected together by kind of graph (network picture). Such models play a crucial role in the design of electronic circuits. They have also been useful in areas like microwave design. But is it ever possible, in principle, to get accurate predictions from this kind of computational model, as we get down to the nanoscale level, where quantum effects like entanglement become crucial to the accuracy of prediction, and where stochastic effects cannot be ignored?

This paper shows that it can be done, for the example of a simple Bell's Theorem experiment[1], in which quantum entanglement is the core of the prediction challenge. Two computational models will be provided, both based on the well-known concept of Markov Random Fields (MRF). As in any lumped-parameter circuit modeling, we first need to develop models of each type of component to be used in the circuit. In simple MRF modeling, the model of each component i is a model of its endogenous probability distribution, $p_i^*(X)$, where X is a possible outcome state of the entire circuit but $p_i^*$ only depends on the state of component i and its immediate neighbors in the graph. Our goal is to predict the probability of outcomes Pr(X) for the circuit as a whole, subject to some kind of boundary conditions (or, more realistically, to predict the total probability of important sets of possible outcomes). We predict Pr(X) in two steps:

$$P^*(X) = p_1(X)p_2(X)p_3(X)...p_n(X) \qquad (1)$$
$$Pr(X) = P^*(X)/Z , \qquad (2)$$

where the "partition function" Z is simply a constant used to make the probabilities add up to 1. (More precisely, Z is the sum or integral of $P^*(X)$ over all possible outcomes X.) We will refer to the quantities $p^*$ and $P^*$ as "relative probabilities."

Research on Bell's Theorem seems to suggest, at first, that MRF models of this kind could not possibly replicate the well-tested predictions of quantum theory for this experiment. The theorems which go with this experiment[2] show that it is impossible for any "local, causal hidden variable" model or simulator to give us the correct predictions. The MRF models to be given here are all local, because each component only interacts with its neighbors. They also meet the definition of "hidden variables" given in the theorem[2], because they are consistent with "realism"[3,4]. However, these calculations do not meet the condition for "causality" as defined in the theorem[2], because they do not march forwards in time from initial conditions to final time. All of the component models are symmetric with respect to time, except for the model of the source of two entangled photons, where forwards-time free energy is being injected from the outside. This results in component models different from the usual time-forwards-only models; however, without such features in the component models, the overall model would be equivalent to the type of classical model which simply cannot give correct predictions, according to the Bell's Theorem and the experiments which go with it[1,2].

---
*The views expressed here were those of the author, not those of his employer.

These simplified models do not qualify as new realistic models of fundamental physics. That would take us beyond the realm of lumped circuit modeling. However, they do raise many interesting questions for future research, to be summarized briefly in the final section of this paper.

## 2 Bell's Theorem experiments with ideal polarizers

### 2.1 Description of the experiment

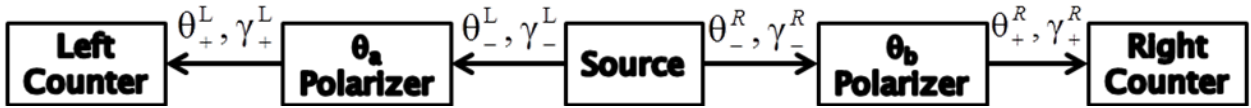

Figure 1. Core structure and notation for the first Bell's Theorem experiments

The core structure of the first Bell's Theorem experiments is shown in Figure 1, which also sets up the notation to be used in this paper. Figure 1 is an abstract version of the design shown in Figure 2 of Clauser and Shimony[1], the most definitive primary source on the original experiments by Clauser, Holt and others. The two polarizers are linear polarizers, tuned to angles $\theta_a$ and $\theta_b$ in the x,y plane. The counters and polarizers are aligned such that they only register photons moving in the direction of ±z, orthogonal to the x,y plane. All these angles are measured "as seen from the source." The source is used to produce pairs of entangled photons, such that $\theta_-^L$ (the polarization angle of the photon emitted on the left channel) equals $\theta_-^R$ (the angle of the photon emitted on the right). In the early experiments, atomic cascades, such as the two-photon decay of excited mercury atoms, were the best available source. Later experiments ruled out local causal hidden variable theories more decisively, using new sources like four-wave mixing and more complex geometry; however, the early experiments are sufficient for our purposes here.

In essence, the task here is to predict the rate at which photons are detected at both counters in coincidence, $R_2(\theta_a-\theta_b)$, as a function of the tuning angles $\theta_a$ and $\theta_b$ which we choose, relative to the rate at which two-photon pairs are produced by the source ($R_0$).

The usual Copenhagen theory of collapse of the wave function predicts that:

$$R_2(\theta_a-\theta_b)/R_0 = \tfrac{1}{2} \cos^2(\theta_a-\theta_b) \qquad (3)$$

Of course, when $\theta_a$ is orthogonal to $\theta_b$, the rate is zero – a prediction which seems very mysterious to many people, and cannot be reconciled with classical ways of thinking. This is the essence of why this is an important experiment.

Actual Bell's Theorem experiments contain additional circuit elements to process the output from the counters, to detect and record coincidence counts, and to filter out noise from uncorrelated photons coming into the experiment. The full analysis also makes allowance for the angular width of the left and right channels; however, Clauser and Shimony report that channels as wide as 30 degrees result in an "F(ξ)" correction of only 1%. This paper will not address the issue of stray photons, and will consider the limiting case where the correction is zero.

This paper will give two MRF models which reproduce the correct quantum mechanical predictions for this experiment. In one model, the transparent model, the outcome array X consists of the six outcome variables which you can see in figure 1: $\theta_-^L$, $\theta_+^L$, $\theta_-^R$, $\theta_+^R$, $\gamma_+^L$, $\gamma_+^R$. The variables $\gamma_+^L$ and $\gamma_+^R$ are just Boolean variables, taking on values 0 or 1, indicating whether or not a photon actually makes it through the left or right polarizer. In another model, the expanded model, we also include four additional Boolean variables, $D_-^L$, $D_+^L$, $D_-^R$, and $D_+^R$, which represent a kind of hidden variable. Intuitively, D = +1 means that the photon is moving forward in time, and D=-1 means that it is moving backwards in time. Mathematically however, the "meaning" of D is simply how we use it to define the MRF model used for prediction.

## 2.2 Quantum mechanical version of the calculation

### 2.2.1 Result of the calculation performed by Clauser and Shimony

Clauser and Shimony report, as their equation 5.9:

$$\left[R(\phi)/R_0\right]_{\Psi_0} = \tfrac{1}{4}(1+\cos 2\phi) \tag{4}$$

Earlier, in section 5.1.2 of Clauser and Shimony[1], they define:

$$\phi = |a - b| \tag{5}$$

where a and b are their notation for $\theta_a$ and $\theta_b$. By elementary trigonometry, this is the same as equation 3 above. They define $R_0$ as the two-photon coincidence counting rate we would expect with the experiment illustrated in Figure 1 if the polarizers were removed; in other words, if the counters record every coincidence, $R_0$ is the rate of production of entangled photons. (Clauser and Shimony do allow for imperfect counters, but the effects of counter imperfection cancel out anyway on the left-hand side of equation 4.) $R(\phi)$ is the rate at which two photons from an entangled pair actually reach the counters. $\Psi_0$ refers to the initial two-photon wave function produced by the source.

### 2.2.2 Derivation

The derivation in Clauser and Shimony (section 5.2.1) is essentially just an application of standard, canonical QED in the Copenhagen version. In this paper, we will refer to that version of QED as KQED. We use "K" for Kopenhagen, rather than C, in order to avoid confusion with cavity or circuit QED, the established meaning of CQED. This section will review their derivation, and highlight aspects which lead into the MRF computation.

In KQED[5,6,7], the wave function $\Psi$ of a system of two photons is a function $\Psi(s_1, \underline{x}^{[1]}, s_2, \underline{x}^{[2]})$ of the spin coordinates $s_i$ and spatial coordinates $\underline{x}^{[i]}$ of the two photons. In their analysis, they do not need to write out the spatial coordinates explicitly, because of the simple geometry. They are allowed to represent the wave function as a function of spin vectors measured as linear polarization, because this experiment does not involve the kind of interactions which require the more usual choice of circular polarization. They do actually write out the "z" component of the spin/polarization vector, but that component is always zero in their calculations; thus, to explain their logic, we can represent the wave function of the two-photon system as a function $\Psi(s_1, s_2)$, where $s_i$ can take on just two values, 1 or 2, representing the x and y coordinates. Clauser and Shimony choose to represent this wave function as a matrix, which I will denote as $\psi$ defined by:

$$\psi = \begin{bmatrix} \Psi(1,1) & \Psi(1,2) \\ \Psi(2,1) & \Psi(2,2) \end{bmatrix} \tag{6}$$

In this notation, equation 5.7 of Clauser and Shimony gives the initial entangled state of the two photons as:

$$\psi^0 = \frac{1}{\sqrt{2}}\left(\underline{e}_1\underline{e}_1^H + \underline{e}_2\underline{e}_2^H\right) = \frac{1}{\sqrt{2}}\left(\begin{bmatrix}1\\0\end{bmatrix}\begin{bmatrix}1 & 0\end{bmatrix} + \begin{bmatrix}0\\1\end{bmatrix}\begin{bmatrix}0 & 1\end{bmatrix}\right) = 2^{-\frac{1}{2}}I \tag{7}$$

In this notation, it is easy to calculate what happens to the wave function matrix ψ if we rotate the polarization of the left-channel photon by an angle ϕ; we get:

$$L(\phi)\psi^0 = \sum_k L_{ik}(\phi)\psi^0_{kj}, \tag{8}$$

where L(ϕ) is the usual rotation matrix of O(2):

$$L(\phi) = \begin{bmatrix} \cos\phi & \sin\phi \\ -\sin\phi & \cos\phi \end{bmatrix} \tag{9}$$

However, the action of a polarizer tuned to an angle θ is more complicated. In KQED, the ideal polarizer is represented as a measurement operator, M(θ), which projects the incoming wave function into the line in spin space which represents the polarization θ. In other words, the incoming wave function is decomposed into two parts, the part along the angle θ, and the part orthogonal to it; the former is just passed through, while the latter is zeroed out. The effect is that the outgoing wave function is all polarized along the angle θ, with a probability amplitude of cos(θ-θ.) times the probability amplitude of the incoming wave. Because actual probabilities are the square of probability amplitudes, this implies that the probability of the incoming photon to be absorbed is:

$$1 - \cos^2(\theta-\theta_\cdot) = \sin^2(\theta-\theta_\cdot) \tag{10}$$

When there are two polarizers in the system, as in Figure 1, where one polarizer applies to the left-channel photon and the other to the right, KQED predicts an outgoing wave function of:

$$\psi^+_{ij} = \sum_{k,m} M_{ik}(\theta_a)\psi^0_{km}M_{mj}(\theta_b), \tag{11}$$

or simply:

$$\psi^+ = M(\theta_a)\psi^0 M(\theta_b) = M(\theta_a)(2^{-\frac{1}{2}}I)M(\theta_b) = 2^{-\frac{1}{2}}M(\theta_a)M(\theta_b) \tag{12}$$

The photon counting rate, per emission of entangled pairs, is simply Tr($\psi^+\psi^{+H}$).

Clauser and Shimony refer to Horne's PhD thesis for the trigonometric calculations which lead from here to equation 2. However, one can see what is happening here by direct inspection. When we apply the operator M($\theta_a$) to the output of M($\theta_b$), as in equation 12, we are applying M($\theta_a$) to a wave function polarized entirely in the direction $\theta_b$. By the very definition of this measurement operator, we know that it reduces the probability amplitude by a factor of cos($\theta_a-\theta_b$), and the probabilities by cos²($\theta_a-\theta_b$), compared to what we would get if $\theta_a=\theta_b$. If we had had $\theta_a=\theta_b$, we would have had effectively only one polarizer; in that case, the probability of a photon getting through, or or being absorbed, would be exactly ½, since our initial incoming wave has equal energy in the $\theta_a$ eigenspace and in the eigenspace of spins orthogonal to $\theta_a$. This leads directly to equation 3.

Notice that the probability amplitude calculation in equation 12 looks exactly like the calculation we would get if a single photon were emitted from the left hand side of figure 1, propagated backwards in time through the left side polarizer, and then through the right-side polarizer after that, except for the factor of $2^{-\frac{1}{2}}$. Klyshko developed this "biphoton" picture in great detail; it has been used extensively by the experimental group of Yanhua Shih[8,9] to design some of the most accurate pioneering experiments in quantum optics. Both of the MRF models here were inspired in great part by that work. In effect, the main contribution of this paper is to translate that intuitive picture into a more general type of computational model.

## 3 How to do the calculation using MRF: The transparent model (MRF1)

### 3.1 Specification of the model

To specify a simple MRF model, following equations 1 and 2, we need only specify the endogenous probabilities, $p_i^*$, for each of the three types of object in Figure 1.

For each of the two counters, our model is:

$$p^*(\gamma_+, \theta_+) = 1 - \gamma_+ + \alpha\gamma_+ d\theta_+ \qquad (13)$$

where $\alpha$ is a small number, on the order of the rate of black body radiation of optical-frequency photons at room temperature (the temperature at which these experiments are normally done). This model reflects the symmetry in time between forwards-time black body emission and the photoelectric effect, explained in Einstein's PhD thesis and classic paper on the photon. In the case where a photon is absorbed by the counter ($\gamma_+=1$), this model gives a uniform distribution across the possible values for its polarization ($\theta_+$).

Our model for the source of two entangled photons is:

$$p^*(\theta_-^L, \theta_+^R) = \delta(\theta_-^L - \theta_+^R) \qquad (14)$$

In order to reproduce the predictions of quantum theory here, without expanding the list of outcome variables, we are forced to use a relatively complex model of the polarizers. For convenience in section 3.2, this model may be written as:

$$p^*(\gamma_+, \theta_+, \gamma_-, \theta_-) = p_A^* + p_B^* + p_C^* + p_D^*,$$

where (15)

$$p_A^* = \delta(\theta_+ - \theta)\delta(\theta_- - \theta) \qquad (16)$$
$$p_B^* = \alpha(\delta(\theta_- - \theta + (\pi/2))\gamma_-(1-\gamma_+) + \delta(\theta_+ - \theta + (\pi/2))\gamma_+(1-\gamma_-)) \qquad (17)$$
$$p_C^* = \alpha(\cos^2(\theta_- - \theta)\delta(\theta - \theta_+) + \cos^2(\theta_+ - \theta)\delta(\theta - \theta_-))\gamma_-\gamma_+ \qquad (18)$$
$$p_D^* = \alpha^2(\sin^2(\theta_- - \theta)\gamma_-(1-\gamma_+) + \sin^2(\theta_+ - \theta)\gamma_+(1-\gamma_-)) \qquad (19)$$

where $\theta$ is the angle ($\theta_a$ or $\theta_b$) which this polarizer is tuned to. The term $\alpha$ multiplying the classical transmission factor $\cos^2(\theta_- - \theta)$ is necessary to avoid the classical prediction that there will be a significant rate of two-photon coincidence even when $\theta_a$ is orthogonal to $\theta_b$. The first term weights the probabilities towards situations like the Klyshko picture, in which $\theta_-$ for one of the two channels equals the tuning angle $\theta$ of one of the polarizers. The second term has a similar effect, in giving us the correct quantum mechanical prediction of the rate at which both photons are absorbed by polarizers. The last two terms preserve the more conventional behavior in Klyshko's picture of what happens when the "biphoton" is moving back forwards in time on the other channel of the experiment[8,9]. A factor of $\alpha$ is added for absorption events over pass-through events, in order to balance the $\alpha$ for absorption at the counters.

### 3.2 Predictions of the model

In the limit as $\alpha \to 0$, this model yields exactly the same predictions as quantum mechanics, as in equation 3 and 4, and also the corresponding prediction from quantum mechanics that both photons will be absorbed, as a function of the angle $\phi = \theta_a - \theta_b$. (For purposes of these calculations, this is equivalent to equation 5, since we only use $\phi$ in the expressions $\sin^2\phi$ and $\cos^2\phi$.) Since $\alpha$ is small at room temperature, terms higher order in $\alpha$ make contributions much smaller than the measurement error in these experiments.

In order to show that the model replicates equation 3, we use the model to calculate the probability that $\gamma_+^L = \gamma_+^R = 1$, conditional upon the production of two entangled photons at the source, i.e. $\gamma_-^L = \gamma_-^R = 1$. In theory, this requires us to calculate the relative probabilities P* for all possible outcomes X for which $\gamma_+^L=\gamma_+^R=1$, integrate them over all four continuous variables in X, and divide by Z. However, in the limit as $\alpha\to 0$, most of these terms are of order $\alpha^4$ or higher, and may be ignored. Also, for simplicity, we will focus on the case where $\theta_a\neq\theta_b$ and $\theta_a$ is not precisely orthogonal to $\theta_b$; the experiments reported in Clauser and Shimony[1] focus on such normal cases.

Under these conditions, the set of possible outcomes X which have nonzero relative probabilities of order $\alpha^3$ is made up of only four subsets:
(1) "Klyshko double coincidence outcomes," where $\theta_-^L=\theta_a$ or $\theta_-^R=\theta_b$ and $\gamma_+^L=\gamma_+^R=1$;
(2) "Klyshko absorption outcomes," where $\theta_-^L=\theta_a$ and $\gamma_+^L=1$ but $\gamma_+^R=0$, or the same with left and right reversed;
(3) antiKlyshko double absorption cases, where $\gamma_+^L=0$ but $\gamma_-^L=1$ and $\theta_-^L=\theta_a\pm(\pi/2)$, and the photon is absorbed in the right such that $\gamma_+^R=0$, or the same with left and right reversed;
(4) antiKlyshko single counting cases, where $\gamma_+^L=0$ but $\gamma_-^L=1$ and $\theta_-^L=\theta_a\pm(\pi/2)$, but $\gamma_+^R=1$, or the same with left and right reversed;

Equations 1 and 2 tell us that the probability of a double coincidence here equals the sum of the relative probabilities for the first subset, divided by Z, which is the sum of the relative probabilities across all four subsets. This is a straightforward calculation.

For Klyshko double coincidence outcomes from the left channel, the product $p_1^*p_2^*p_3^*p_4^*p_5^*$ in equation 1 becomes:

$$(\alpha d\theta_+^L)p_A^*\delta(\theta_-^L-\theta_-^R)p_C^*(\alpha d\theta_+^R) \tag{20}$$

When we integrate this over the angles $\theta_-$, $\theta_-^R$ and so on, the delta functions disappear, giving us a simple result:
$$\alpha(\alpha\cos^2\phi)\alpha \tag{21}$$
The total over both channels, left and right, is simply:
$$2\alpha(\alpha\cos^2\phi)\alpha \tag{22}$$
Likewise, for Klyshko absorption outcomes, the product $p_1^*p_2^*p_3^*p_4^*p_5^*$ in equation 1 becomes:
$$(\alpha d\theta_+^L)p_A^*\delta(\theta_-^L-\theta_-^R)p_D^*, \tag{23}$$
which, when integrated and summed over both channels, becomes:
$$2\alpha(\alpha\sin^2\phi)\alpha \tag{24}$$
For antiKlyshko double absorption cases, the product $p_1^*p_2^*p_3^*p_4^*p_5^*$ in equation 1 becomes:
$$p_B^*\delta(\theta_-^L-\theta_-^R)p_D^*, \tag{25}$$
which (considering how a change in angle of $\pm\pi/2$ changes $\sin^2$ into $\cos^2$) yields an integral summed over the two channels of:
$$2\alpha(\alpha^2\cos^2\phi) \tag{26}$$
Finally, the antiKlysho partial absorption cases have relative probabilities of
$$p_B^{**}\delta(\theta_-^L-\theta_-^R)p_C^*(\alpha d\theta_+^R) \tag{27}$$
which integrate and sum to a total relative probability for that subset of:
$$2\alpha(\alpha\sin^2\phi)\alpha \tag{28}$$
The total partition function Z is the sum of the total relative probabilities across all four subsets of X, as given in equations 22, 24, 26 and 28:

$$Z = 2\alpha(\alpha\cos^2\phi)\alpha + 2\alpha(\alpha\sin^2\phi)\alpha + 2\alpha(\alpha^2\cos^2\phi) + 2\alpha(\alpha\sin^2\phi)\alpha = 4\alpha^3 \tag{29}$$

Following equation 2, the probability of a double coincidence is the total relative probability of such outcomes, as given in equation 22, divided by $Z=4\alpha^3$; that results directly in the prediction of equation 3.

# 4 The expanded model: a simpler alternative (MRF2)

## 4.1 Specification of the model

Again, we specify the model simply by specifying the component models for each of the 3 types of object in Figure 1.
For the counters, we still use equation 13. In this case, since there is an additional outcome variable $D_+$ coming into the counter, equation 13 gives us a probability distribution which is uniform both in $\theta_+$ and in $D_+$. In other words, it gives equal probability to the possibility that $D_+=-1$ as to $D_+=+1$.

For the source of entangled photons, we now have:

$$p^*(\theta_-^L, \theta_+^R, D_-^L, D_-^R) = \delta(\theta_-^L - \theta_+^R)(1-D_-^L D_-^R) \tag{30}$$

which is like equation 14, except that it enforces the concept that the two photons must travel "in opposite directions in time." Notice that we could have divided this expression by 2, for esthetic reasons, but that would not affect the calculations of Pr(x) at all.

The benefit of adding the D variables is that we can get correct predictions using a much simpler model of the polarizers in the experiment:

$$p^*(\gamma_+, \theta_+, D, \gamma_-, \theta_-) = (\cos^2(\theta_- - \theta)(1+D) + \cos^2(\theta_+ - \theta)(1-D))\gamma_-\gamma_+$$
$$+\alpha(\sin^2(\theta_- - \theta)(1+D)\gamma_-(1-\gamma_+) + \sin^2(\theta_+ - \theta)(1-D)(1-\gamma_-)\gamma_+)) \tag{31}$$

where $p^*$ is nonzero only when $D_-=D_+=D$. The presence of the D variables avoids the presence of spurious classical terms, and makes it possible to avoid the complexity of extra terms to dial down their magnitude.

## 4.2. Predictions of the model

The equivalence to quantum mechanics is much more obvious here than in the transparent model. Equation 30 enforces the concept that relative probabilities are zero, except in cases where a photon is moving "backwards in time" from one of the channels (e.g. $D_-^L=-11$, if it comes from the left) and then its partner moves "forwards in time" down the other channel. Equations 30 and 31 lead exactly to the same probabilities as we would get for a single photon moving forward in time through two polarizers – except for the antiKlyshko cases. Here, as in section 3.2, the total relative probability of the antiKlyshko cases is exactly the same as that of the Klyshko cases, which has the effect of doubling Z, and dividing the rate of coincidence in half.

Still, let us go back and consider the details.

In calculating the total relative probability for a double coincidence case, with $D_-^L=1$, we need to remember to only count cases where $\gamma_-^L=1$, meeting the boundary conditions for the conditional probability we are trying to compute. From the polarizer model in equation 31, we know that this requires that $\theta_-^L=\theta_a$. We also know that half the original energy or probability gets lost when the initial angle $\theta_+^L$ comes from a uniform distribution. And of course, probability is zero unless $\theta_-^R=\theta_-^L = \theta_b$. Putting this together we arrive at a total relative probability of double coincidences "from the left" of

$$(1/2)\alpha(\cos^2\phi)\alpha \tag{32}$$

The sum across left and right channels is:

$$\alpha^2\cos^2\phi \tag{33}$$

Likewise, the total relative probability for a Klyshko absorption event is similar, except that there is absorption on the channel where the partner photon goes forwards in time, yielding a sum across both channels of:

$$\alpha(\alpha \sin^2\phi) \tag{34}$$

As in section 3.2, the total relative probability of the antiKlyshko events is just the same as that of the Klyshko outcomes, such that $Z=2\alpha^2$. Dividing equation 32 by $Z=2\alpha^2$, we again arrive at equation 3.

# 5. Implications and possibilities for future research

## 5.1. Immediate questions

In previous sections, we have carefully refrained from making editorial remarks about the meaning of these results, for two reasons. First, the results raise interesting questions in their own right, and it is better that readers think about these questions from whatever viewpoint they find most interesting. Second, there are many different ways that one might follow up here.

      One might follow up by applying this MRF approach to other quantum optics circuits. There are many other systems which can be represented by a graph like Figure 1, and a few new component models added to the stable of three models provided here (both for transparent and expanded modeling). For example, Klyshko's former collaborators have analyzed and experimented with a kind of "triphoton" extension of their earlier work on Bell's Theorem[10]. Or perhaps other systems from the same group might be easier to handle in lumped parameter modeling. After more component and examples are developed, it may be worthwhile to automate the process of working out the probabilities, by using mathematical tools like the MRF automated inference methods developed in artificial intelligence or image processing. Perhaps someday such tools might even be used to help design more complex circuits which themselves could be used to model or simulate even more complex systems.

      Another way to follow up would be to probe more deeply into polarizers themselves, and address some of the unanswered questions in the discussion of Clauser and Shimony[1]. Clauser and Shimony address the issue of how to model imperfect polarizers, still at a macroscopic level. They implicitly assume a new measurement model $M'(\theta)$ for imperfect polarizers. However, both in KQED and in MRF modeling, there are other equally plausible ways that one might model imperfections in polarizers, such as the calcite prism polarizers used in the experiments of Holt. Perhaps different models might work for different polarizers, and help explain some of the variations in results in these experiments. Or perhaps not. Empirical methods have been developed to characterize such systems more systematically; however, so far as we know, they have only been used to validate larger systems[11], and not to probe the exact nature of polarizer imperfections. The polarizer models here also suggest that study of the black body property of polarizers, such as red hot calcite prisms, might yield interesting information. In the long term, of course, it is important to connect this kind of phenomenological input-output modeling of polarizers to the condensed matter physics of such objects.

      Beyond these obvious directions, we are intrigued by the sheer elegance of the expanded model versus the transparent model. Could it be that we have been missing an important degree of freedom when we quantize the photon, a degree of freedom which usually can be ignored but not always? We have not yet had time to think hard about this question.

      The initial motivation for this work actually came from two other lines of research – an investigation of axiomatic quantum field theory, and the search for computational methods to cope with more complex electrooptic systems where today's modeling tools appear to have fundamental weaknesses.

      For axiomatic quantum field theory, Robert O'Connell pointed us towards work which suggests that the usual methods of KQED are inherently ambiguous, and that we need work to try to figure out how to use more well-defined axiomatic versions of QED in practical calculations[12]. MRF methods, in the form of lattice dynamics, have been widely used in some of the axiomatic work[13,14,15].

      But for more general classes of system, lumped parameter modeling and realistic MRF methods may have their own limits. Equations 1 and 2 implicitly assume that each component in a circuit has its own probability distribution, which operates independently. They are analogous to Gibbsian thermodynamics, which breaks down when we try to model the tightly coupled interaction of atoms and electrons within a crystal such as a semiconductor. A new Boltzmann-based approach may be needed, in order to push three dimensional computational modeling as far as it can go into the quantum domain. Energy levels and density of states may be the "glue" to integrate the components, rather than probability as such; in other words, a more general way to model $Pr(X)$ is by evaluating $Pr(-kH)dx$, where $H$ and $dx$ play the decisive roles. However, new sophisticated methods of that sort must be able to cope with simple examples,

like the example discussed in this paper. Our real motivation to do this work is to understand what this requires, not only for MRF methods but for more sophisticated and complex methods in the same family.

**5.2. Possible answers and all-angles triphoton experiment**

After this paper was written, I have made some efforts to begin to answer some of the questions here. In a new paper[16], I develop some new tools, to help generalize the MRF approach across space-time to the case of continuous fields. In a kind of sequel[17] to this paper, I provide yet a third MRF model of the Bell's system experiment, closer to the actual physics, and discuss how to marry up this approach to something like Carmichaels' Quantum Trajectory approach.

On closer examination, it is still unclear as yet whether the predictions of any local MRF model *will* disagree in some cases with those of the standard Copenhagen form of quantum mechanics, in which measurement operators immediately change the wave function of density matrix of the entire universe. More precisely, there is a question whether these models would disagree with the traditional version of quantum mechanics in which the measurement operator is modeled as the linear mapping defined by:

$$M(\theta_b): |\psi><\psi| \Rightarrow c_b |\psi_b><\psi_b| + c_b^- |\psi_b+\pi/2><\psi_b+\pi/2| \quad (35)$$

where $c_b=(<\theta_b|\rho(t-\varepsilon)|\theta_b>)$ and $c_b^- = 1-c_b$, where $\psi_b$ is the projection of the wave function $\psi$ to points in Fock space for which the polarization of photon b is $\theta_b$ (the angle of the polarizer it reaches), and where "$\psi_b+\pi/2$" is the projection of the wave function $\psi$ to points for which the polarization of b is $\theta_b+\pi/2$. However, *this disagreement would occur in experiments which have not yet been performed*. Of special interest is the all-angles triphoton experiment illustrated in Figure 2:

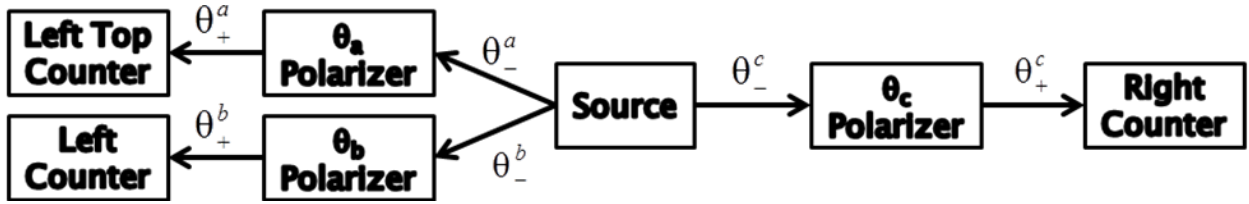

Figure 2. A possible triphoton experiment (taken from the sequel[17]).

As a source, we could use the triphoton source implemented in experiments in Zeilinger's group. From equation 4 of their paper[18], it would seem possible to implement a source which outputs the wave function:

$$\psi = \frac{1}{\sqrt{2}}\left[|0\rangle_a|0\rangle_b\left|\tfrac{\pi}{2}\right\rangle_c + \left|\tfrac{\pi}{2}\right\rangle_a\left|\tfrac{\pi}{2}\right\rangle_b|0\rangle_c\right] \quad (36),$$

where I use "0" to represent a state of linear polarization in the horizontal direction, and "$\pi/2$" to represent polarization in the orthogonal direction (vertical). The Appendix discusses in detail how this experiment would be decisive, in allowing us to test whether the traditional model of wave function collapse in the polarizer is correct, or whether the new type of time-symmetric model is (unless of course the results call for a third type of model). It briefly states what the implications would be for the correct formulation of QED if the time-symmetric version[4] wins out.

# Appendix. All-Angles Triphoton Experiment As Test of Collapse of the Wave Function

This appendix calculates what the three-photon counting rate is predicted to be in general, for the all-angles triphoton experiment, as a function of all possible polarizer angles, according to the "collapse of the wave function" model of the polarizers which was previously successful in predictions of Bell's Theorem experiments by Clauser and his collaborators. No metaphysical observers are assumed in this model. A simple master equation is introduced which represents the condensation of the wave function as a physical effect within the polarizer. The final section also reviews a corrected version of what time-symmetric physics predicts for the same situation[22], and gives a table showing that the predictions are decisively different.

## A.1 Introduction and Summary

The purpose of this appendix is to calculate the predictions of the outcome of the proposed experiment using the same quantum mechanical calculation method used in previous successful predictions of Bell's Theorem (biphoton) experiments, as discussed in section 2.2 above.

The original, simple Bell's Theorem experiments were as shown in Figure 1. In the limit of ideal polarizers, the original successful prediction was:

$$R_2/R_0 = \tfrac{1}{2} \cos^2(\theta_a - \theta_b), \tag{37}$$

where $R_2$ is the rate of detection of two coincident photons at the detectors and $R_0$ is the rate of production of two entangled photons at the source. The purpose of this appendix is to calculate predictions for a similar counting rate, the three-particle counting rate for the triphoton experiment shown in Figure 2 and equation 36 above. Here we will consider only the case where the polarizers are aligned such that the "a" photon goes through (or is absorbed by) the $\theta_a$ polarizer, after which the "b" photon goes through (or is absorbed by) the $\theta_b$ polarizer, after which the "c" photon reaches the "c" polarizer. As before, this will be done in the case of ideal polarizers and detectors.

The successful predictions by Clauser and others of the Bell's Theorem experiment were based on a "collapse of the wave function" model of the polarizers. Applying that same model here, we will show that we get the prediction:

$$R_3/R_0 = 1/2\, (\cos\theta_a \cos\theta_b \sin\theta_c + \sin\theta_a \sin\theta_b \cos\theta_c)^2. \tag{38}$$

## A.2. Collapse of the Wave Function

Many, many formulations of quantum mechanics have appeared over the past few decades, not only in physics but in mathematics and in philosophy. This paper will not address the question of which version is equivalent to which. It will focus only on the specific early Copenhagen version, used in previous successful predictions, in which the polarizers are modeled as agents of a time-asymmetric "collapse of the wave function."

More precisely, when one of the photons described in a wave function $\psi$ reaches a linear polarizer, it is assumed that the wave function is immediately decomposed (projected) into two parts, the part in which that photon has the linear polarization preferred by the polarizer, and the part in which it has the orthogonal polarization. The first component is passed through the polarizer, and the second component is absorbed. The macroscopic outcome is stochastic. With probability equal to the square length of the first component, the first component will emerge out of the polarizer, now normalized to a length of one. With probability equal to the square length of the other component, nothing emerges from the polarizer.

In modern notation, such stochastic outcomes may be described in terms of density matrices $\rho$, widely used in quantum optics, which can represent statistical ensembles of possible wave functions, such as:

$$\rho = \sum_{k=1}^{n} \Pr(k)\psi_k \psi_k^H \tag{39}$$

where, following Schwinger's notation, I use a superscript "H" to denote the Hermitian conjugate. Density matrices have been widely used in condensed matter physics and in quantum optics, where they are essential in modeling the impact of random effects in the "reservoirs" of solid matter which light passes through in realistic experiments[19,20]. Past calculations of Bell's Theorem predictions have modeled the polarizer as a bulk object, performing condensation of the wave function, but an equivalent result follows if we model the polarizer in continuous time, as an object which implements the simple dynamics:

$$\dot{\rho} = g a(\theta_p + \tfrac{\pi}{2}) \rho a^+(\theta_p + \tfrac{\pi}{2}) \tag{40}$$

between times $t_-$ and $t_+$, where $t_-$ is the time when a photon enters the polarizer and $t_+$ is when it leaves (if not absorbed), where g is a parameter, and where $a(\theta_p+\pi/2)$ is the annihilation operator for a photon of linear polarization $\theta_p+\pi/2$ in the polarizer. For g and $t_+-t_-$ large enough, we approach the case of an ideal polarizer. Here, I write $\theta_p$ as the preferred angle of linear polarization to pass through this polarizer, such that $\theta_p+\pi/2$ is the orthogonal polarization which it absorbs. I will refer to this simple continuous-time model of a certain type of polarizer as CQMp in future papers.

CMRFp would not be a good model of calcite type polarizers[17], which rely heavily on internal reflection, but it seems reasonable enough for the polaroid type of polarizers which have also been widely used in these experiments. This is a simple example of a master equation[19,20], assuming that stochastic effects and noise are caused by events within the polarizer itself, and not by phenomena such as quantum interference between graduate students examining the printouts of the results of the experiments, acting as metaphysical observers; some philosophers might prefer the latter type of model, but we are not aware of how to formulate such models in a way which would yield quantitative predictions here.

The remainder of this paper will make no further reference either to master equations or to metaphysical observers. It will simply apply the collapse-of-the-wave function model of the polarizer, as in the work of Clauser et al, to the task of predicting the new experiment.

### A.3. Calculation of the Prediction

To apply the traditional "collapse of the wave function" model, we begin by reviewing the wave function in effect when light reaches the first polarizer, assumed to be the $\theta_a$ polarizer:

$$\psi = \frac{1}{\sqrt{2}}\left[|0\rangle_a|0\rangle_b\left|\tfrac{\pi}{2}\right\rangle_c + \left|\tfrac{\pi}{2}\right\rangle_a\left|\tfrac{\pi}{2}\right\rangle_b|0\rangle_c\right] \tag{41},$$

From elementary trigonometry, we know that the following holds for any angle $\theta$ (see figure 3):

$$|0\rangle = \cos\theta|\theta\rangle - \sin\theta|\theta + \tfrac{\pi}{2}\rangle \tag{42}$$
$$\left|\tfrac{\pi}{2}\right\rangle = \sin\theta|\theta\rangle + \cos\theta|\theta + \tfrac{\pi}{2}\rangle \tag{43}$$

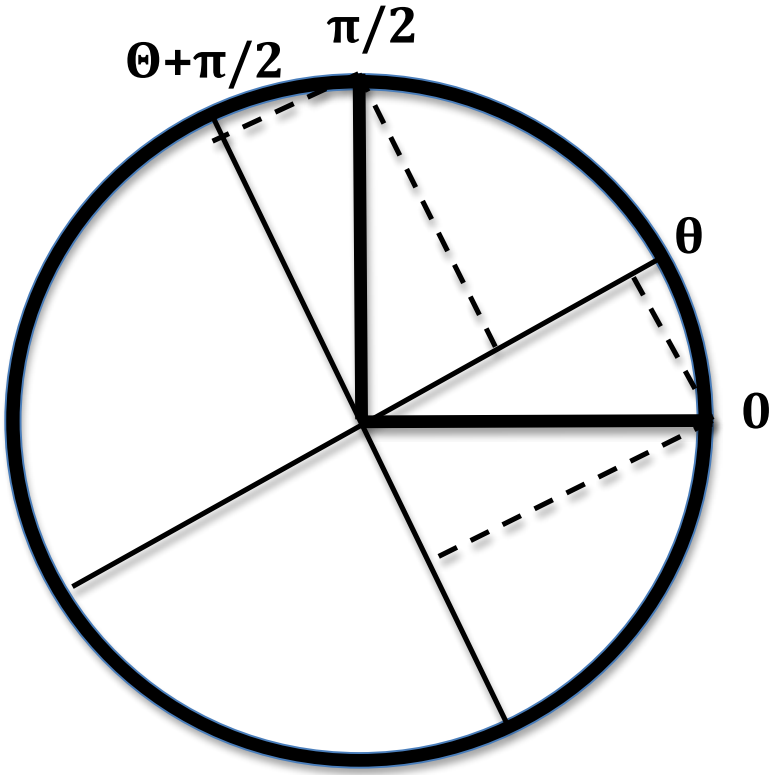

Figure 3. Geometric picture of equations 42 and 43

If we consider the case where $\theta=\theta_a$, and we define $c_a$ as $\cos\theta_a$ and $s_a$ as $\sin\theta_a$, equations 42 and 43 tell us that:

$$|0\rangle_a = c_a|\theta_a\rangle_a - s_a|\theta_a + \tfrac{\pi}{2}\rangle_a \tag{44}$$

$$\left|\tfrac{\pi}{2}\right\rangle_a = s_a \left|\theta_a\right\rangle_a + c_a \left|\theta_a + \tfrac{\pi}{2}\right\rangle_a \tag{45}$$

By substituting equations 44 and 45 into equation 41, we may deduce that:

$$\psi = \frac{1}{\sqrt{2}}\left[\left|\theta\right\rangle_a \psi^+ + \left|\theta_a + \tfrac{\pi}{2}\right\rangle_a \psi^-\right] \tag{46}$$

where we define:

$$\psi^+ = c_a \left|0\right\rangle_b \left|\tfrac{\pi}{2}\right\rangle_c + s_a \left|\tfrac{\pi}{2}\right\rangle_b \left|0\right\rangle_c \tag{47}$$

$$\psi^- = -s_a \left|0\right\rangle_b \left|\tfrac{\pi}{2}\right\rangle_c + c_a \left|\tfrac{\pi}{2}\right\rangle_b \left|0\right\rangle_c \tag{48}$$

Of course, equation 46 is the decomposition of the oncoming wave function into two components, as called for in the collapse-of-the-wave-function model. Since $\psi^+$ and $\psi^-$ both have a square length of 1, the model predicts a probability of ½ that a photon will be detected at a, and it predicts that $\psi^+$ is the active component of the outgoing wave function which will be sent to polarizers $\theta_b$ and $\theta_c$ in the case where there will be a detection on the "a" channel. This tells us that $R_3/R_0$ is predicted to equal ½ of the predicted rate $R_2/R_0$ for the Bell-like experiment in which the wave function $\psi^+$ is propagated to the polarizers $\theta_b$ and $\theta_c$.

To predict what happens next, when the "b" photon arrives at the $\theta_b$ polarizer, we again exploit equations 42 and 43 to deduce:

$$\left|0\right\rangle_b = c_b \left|\theta_b\right\rangle_b - s_b \left|\theta_b + \tfrac{\pi}{2}\right\rangle_b \tag{49}$$

$$\left|\tfrac{\pi}{2}\right\rangle_b = s_b \left|\theta_b\right\rangle_b + c_b \left|\theta_b + \tfrac{\pi}{2}\right\rangle_b \tag{50}$$

Substituting equations 49 and 50 into equation 47, we deduce that:

$$\psi^+ = \left|\theta_b\right\rangle_b \left[c_a c_b \left|\tfrac{\pi}{2}\right\rangle_c + s_a s_b \left|0\right\rangle_c\right] + \left|\theta_b + \tfrac{\pi}{2}\right\rangle_b \left[-c_a s_b \left|\tfrac{\pi}{2}\right\rangle + s_a c_b \left|0\right\rangle_c\right] \tag{51}$$

As before, only the left side contributes to the three-photon detection rate. According to the collapse of the wave function, the probability of photon b emerging from polarizer $\theta_b$ is the square norm of the wave function on the left side, which we may denote as:

$$C_{ab} = c_a^2 c_b^2 + s_a^2 s_b^2 \tag{52}$$

Since this is a new probability factor, assuming we already have a photon "a" emerging from $\theta_a$, we are in the following situation. With probability $(1/2)C_{ab}$, photons a and b both will emerge (and be detected), and the following normalized wave function goes to the final polarizer, $\theta_c$:

$$\psi^{++} = \frac{1}{\sqrt{C_{ab}}}\left[c_a c_b \left|\tfrac{\pi}{2}\right\rangle_c + s_a s_b \left|0\right\rangle_c\right] \tag{53}$$

Using the same sort of substitution as before, we easily see that the probability of photon c making it through the third polarizer is predicted to be:

$$\left(\frac{1}{\sqrt{C_{ab}}}(c_a c_b s_c + s_a s_b c_c)\right)^2 \tag{54}$$

Multiplying this by $(1/2)C_{ab}$, we directly arrive at equation 38 for the overall three-photon coincidence rate.

## A.4. Decisive Difference From the MRF (Time-Symmetric) Case and Broader Significance

Because the MRF models embody the intuitive insights of Klyshko[9,10], it is possible that earlier work inspired by Klyshko would also make predictions different from those which result from assuming collapse of the wave function. In 2015, I published a paper[22] arguing that the MRF models (and time-symmetric physics in general as I defined it in 2008[4]) would predict a three-photon counting rate of:

$$R_3/R_0 = k \cos^2(\theta_c - \theta_b - \theta_a) \quad (55),$$

**if** we assume that the source is proportional to $\delta(\theta_c - \theta_a - \theta_b)$, similar to equation 14. However, equation 41 clearly adds a factor of $\pm\pi/2$ to $\theta_c$, such that a corrected prediction would be:

$$R_3/R_0 = k \sin^2(\theta_c - \theta_b - \theta_a) \quad (56),$$

which clearly implies predictions quite different from equation 38, as can easily be verified on a simple spreadsheet.

In my view, the experiment will probably support equation 56 over equation 38, because the original collapse model was relatively ad hoc, while the requirement for time-symmetry in modeling passive objects like polarizers follows from a very fundamental analysis of how measurement works in a multiverse whose dynamics are governed by a time-symmetric Schrodinger equation[4]. However, until the experiment is actually done, we will not really know. A preliminary effort was done to do this experiment using a source obeying equation 41, using thermal light[21], but Shih reported in a conference in Princeton in 2015 that the results did not agree either with equation 38 or equation 55, and that the thermal light source was probably not producing the true asymmetric GHZ state (equation 41).

If a more definite experiment is performed using Down conversion[18] or other reliable GHZ sources[23], and if it agrees with equation 56, this would of course require some change in how we formulate Quantum Electrodynamics (MQED). The most obvious reformulation, Markovian or modified QED, would entail assuming the same "Schrodinger equation" as in canonical QED, but requiring new models of all explicitly modeled passive macroscopic objects like polarizers used in measurement and in other parts of an experiment, conforming with time-symmetric physics[4]. In other words, MQED asks us to replace the existing time-asymmetric measurement models with their symmetrized versions, as this paper has already shown how to do in the important case of polarizers. The implications would go far beyond measurement as such, since they would apply to passive macroscopic objects like spin gates and even transistors used in complex circuits, thereby giving some new degrees of freedom in design. Only the points of injection of forwards-time free energy would be exempt from a need to change the models of macroscopic objects, in principle, though of course the change in models is important only for some designs.